\documentstyle{article}
\begin{document}
\openup6pt

\title
{$\omega$ dependence of the scalar field in Brans-Dicke theory }
\author{A. Bhadra\thanks{e-mail: bhadra@nbu.ernet.in} \\
High Energy and Cosmic Ray Research Centre\\
University of North Bengal, Darjeeling (W.B.) 734 430
INDIA\\
and \\
K.K. Nandi\thanks{ e-mail: kamalnandi@hotmail.com} \\
Department of Mathematics\\
University of North Bengal, Darjeeling (W.B.) 734 430
INDIA
}
\date{}
\maketitle

\begin{abstract}
This article examines the claim that the Brans-Dicke scalar field
$\phi \rightarrow \phi_{0} + $ O$\left( \frac {1}{\sqrt {\omega}
}\right)$ for large $\omega$ when the matter field is traceless.
It is argued that such a claim can not be true in general.
\end{abstract}

PACS numbers: 04.50. +h

Brans-Dicke (BD) theory [1] is generally regarded as a viable
alternative to Einstein's theory of General Relativity. This
theory has recently regained interest because, in the Einstein
conformal frame, it turns out to be the low energy limit of many
theories of quantum gravity such as the supersymmetric string
theory [2] or Kaluza-Klein theory [3]. The theory is relevant also
in the extended inflationay scenario of cosmology [4]. The BD
theory, which accommodates Mach's principle, describes gravitation
through a spacetime metric ($g_{\mu\nu}$) and a massless scalar
field ($\phi$) that couples to both matter and spacetime geometry,
The strength of the coupling is represented by a single
dimensionless constant $\omega$. In the Jordan conformal frame,
the BD action takes the form

\begin{equation}
{\cal A}= \frac{1}{16 \pi }\int d^{4}x \sqrt{-g}\left(\phi R-
\frac{\omega }{\phi }
g^{\mu\nu} \phi_{,\mu} \phi_{,\nu} + {\cal L}_{matter} \right)
\end{equation}

where \( {\cal L} _{matter} \) is the Lagrangian density of
ordinary matter. Variation of (1) with respect to $ g^ {\mu \nu} $
and $\phi$ gives, respectively, the field equations

\begin{equation}
R_{\mu\nu} -\frac{1}{2}g_{\mu\nu}R= \frac{8 \pi} {\phi } T_{\mu \nu}
+\frac {\omega}{\phi ^{2}}\left( \phi_{,\mu} \phi_{,\nu}- g_{\mu\nu}
\phi^{,\sigma} \phi_{,\sigma} \right) +
\frac{1}{\phi} \left( \nabla_{\mu}\nabla_{\nu}\phi-g_{\mu \nu}
 \Box \phi \right),
\end{equation}

\begin{equation}
\Box \phi = \frac {8\pi T}{(2\omega + 3)}
\end{equation}

where $R$ is the Ricci scalar, and $T$=$T_{\mu}^{\mu}$ is the trace
of the matter energy momentum tensor.

In the weak field approximation, the metric tensor can be written as
\begin{displaymath}
 g_{\mu \nu} = \eta_{\mu \nu} + h_{\mu \nu}
\end{displaymath}
where $\eta_{\mu \nu}$ is the Minkowskian metric tensor. Similarly
$\phi$ = $\phi_{0}$ + $\xi$, where $\phi_{0}$ is a constant. Using
these approximations in equatons (2) and (3), one concludes that
[5]
\begin{equation}
\phi = \phi_{0} + O \left( \frac {1}{\omega} \right)
\end{equation}
and
\begin{equation}
R \sim O\left(\frac{1}{\omega}\right).
\end{equation}
Thus it appears from above equations that the post-Newtonian
expansion of BD theory reduces to general relativity in the
infinite $\omega$ limit. But it was reported [6] that a number of
exact solutions of BD theory do not go over to the corresponding
solutions of general relativity in the limit $\omega \rightarrow
\infty$. Recently, Banerjee and Sen [7] illuminated this point
through the study of the BD field equations and pointed out that,
when the trace ($T$) of the energy momentum tensor vanishes, the
asymtotic behavior of $\phi$ is not represented by equation (4)
but follows the relation
\begin{equation}
\phi = \phi_{0} + O\left(\frac{1}{\sqrt {\omega} }\right)
\end{equation}
Faraoni [8] also claimed to have found a similar $\omega$
dependence. As a result, the BD theory does not tend to general
relativity in the $\omega \rightarrow \infty$ limit. This feature
is significant because the lower limit of $\omega (\sim 500)$ for
the solar system measurements is fixed using the
$O\left(\frac{1}{\omega}\right)$ behavior in the standard PPN
approximation. It is therefore important to study the situation
more closely, which we do here.
\\
We noticed that equation (6) is not valid in general. In this
brief report we will discuss some counter examples to equation (6)
in BD theory. We will also point out some assumptions inherent in
[7] and
[8] that led to equation (6).\\
When $T$=0, BD field equations yield
\begin{equation}
\Box\phi = 0
\end{equation}
and
\begin{equation}
R= \frac{\omega}{\phi ^{2}}\left( \phi^{,\alpha}
\phi_{,\alpha} \right)
\end{equation}
From the above equation, Banerjee and Sen [7] argued that $\phi$
will exhibit the asymtotic behavior as given in equation (6). {\it
But such a conclusion holds only if $R$ is assumed to be
independent of} $\omega$. This is a strong condition which is not
justified in general. Note that, equation (8) contains two unknown
funtions of $\omega$: $R$ and $\phi$. Hence if one knows the
dependence on $\omega$ of one of the functions, the same for the
other could be obtained from equation (8). It is true that the
$\omega$ independace of $R$ leads to equation (6) but there is no
way to know the functional behavior of $R$ {\it a priori} unless
one considers specific solutions. On the other hand, it is known
that for a number of exact solutions of BD theory having traceless
source,
$R$ {\it is} a function of $\omega$. \\
To clarify the situation further, let us consider the static
spherically symmeteric vacuum solution of the BD theory given by
Brans and Dicke [1]:
\begin{equation}
ds^{2}= e^{2\alpha} dt^{2} - e^{2\beta} \left( dr^{2} + r^{2}
d\theta ^{2} + r^{2} sin^{2} \theta d\phi ^{2} \right)
\end{equation}
where
\begin{equation}
 e^{2\alpha} = \left( \frac{1-B/r}{1+B/r} \right)^{\frac{2}{\lambda}}
\end{equation}
\begin{equation}
 e^{2\beta} = \left( 1 + \frac{B}{r}\right) ^{4} \left( \frac{1-B/r}
 {1+B/r} \right) ^{\frac{2(\lambda -C -1)}{\lambda}}
\end{equation}
\begin{equation}
 \phi = \phi_{0} \left( \frac{1-B/r}{1+B/r} \right)^{\frac{C}{\lambda}}
\end{equation}
with
\begin{equation}
\lambda = \left( [C+1]^{2} - C[1-\frac{\omega C}{2}]
\right)^{\frac{1}{2} }
\end{equation}
The Ricci scalar for the above metric is given by
\begin{equation}
R =  \frac {4 \omega C^{2} B^{2} r^{4} }{\lambda ^{2} } \left( r+ B \right)
                              ^ {-4 \left[ 1+ \frac {C+1} {\lambda}\right]}
\left( r- B \right) ^ {-4 \left[ 1- \frac {C+1} {\lambda}\right]}
\end{equation}
To establish their claim, Banerjee and Sen leave $C$ to be
arbitrary but it is not clear to what extent it is so. Any choice
of $C$ arbitrarily dependent on $\omega$ will not render $R$ to be
$\omega$ independent.  Only when either $C$ is an arbitrary but
fixed constant or $C(\omega) \propto \frac {1} {\sqrt {\omega}}$
for large $\omega$, does  $R$ become effectively (but not exactly)
independent of $\omega$. Therefore, the arbitrariness of C as used
by Banerjee and Sen is severely constrained. On the other hand, it
is well known that, to match the BD class I metric [1] with weak
field post Newtonian expansion of the BD field equations (which is
a standard and probably unique way to fix unknown constants
present in vacuum solutions), one must specify $C$ = $ \frac{-1}
{2 + \omega} $. And under this choice $R$ goes as $ O\left(
\frac{1}{\omega} \right) $ and so does $\phi$. There are other
examples too. For instance, consider the stationary charged black
hole solutions in BD theory recently obtained by Kim [9] and is
given by
\begin{displaymath}
ds^{2} = \Delta^{ -\frac {2}{2\omega+3}} sin^{ -\frac {4}{2\omega+3}}
 \theta [ -\left(\frac {\Delta - a^{2}sin^{2} \theta}{\Sigma}
 \right)dt^{2} -\frac {2asin^{2} \theta (r^{2}+a^{2}-\Delta)}
 {\Sigma} dt d\phi
\end{displaymath}
\begin{equation}
+ \left( \frac {(r^{2} + a^{2})^{2} -\Delta a^{2}sin^{2}\theta}
{\Sigma} \right) sin^{2}\theta d\phi^{2} ] + \Delta^{ \frac
{2}{2\omega+3}} sin^{ \frac {4}{2\omega+3}}\theta \left( \frac
{\Sigma } {\Delta} dr^{2}+ \Sigma d\theta^{2} \right)
\end{equation}
\begin{equation}
\Phi(r,\theta) = \Delta ^{ \frac {2}{2\omega+3}}
sin^{ \frac {4}{2\omega+3}} \theta
\end{equation}
\begin{equation}
A_{\mu} = \frac {er}{\Sigma} \left( \delta^{t}_{\mu}
- a sin^{2} \theta \delta^{\phi}_{\mu} \right)
\end{equation}
where $ \Sigma = r^{2} + a^{2}cos^{2}\theta $ and $\Delta = r^{2}-
2Mr + a^{2} + e^{2}$ with $M, a$ and $e$ representing the ADM
mass, angular momentum per unit mass and electric charge
respectively. In this case the source is electromagnetic field and
hence traceless. The above solution reduces to the standard
Kerr-Newman solution [10] in the limit $\omega \rightarrow
\infty$. The curvature scalar for the metric is given by
\begin{equation}
R = \omega \left( \frac {4}{2\omega + 3} \right) ^{2} \frac
{1}{\Sigma} sin^ { \frac {-4}{2\omega+3} } \theta \left[ (r-M)^{2}
\Delta^ {-(2 \omega +5)/(2 \omega +3)} + cot^{2}  \theta \Delta^
{-2/(2\omega +3)} \right]
\end{equation}
It is evident from equations (16) and (18) that both the scalar
field and scalar curvature go as O$( \frac {1} {\omega} )$
contradicting equation (6).\\
Faraoni [8] claimed to have deduced a rigorous behavior of $\phi$
in terms of $\omega$ supporting the result of [7]. But the above
examples (BD class I and Kim's black hole solution) already
contradict such a claim. The author [8] used the conformal
invariance of BD theory under the transformations
\begin{eqnarray}
\tilde{g} _{\mu \nu} = \phi ^{2 \alpha} g_{\mu \nu} \\
\tilde {\phi} = \phi ^{1-2\alpha} \\
\tilde {\omega} = \frac {\omega -6 \alpha (\alpha -1)}
{(1-2 \alpha)^{2} },
\end{eqnarray}
$\alpha \neq \frac{1}{2} $.
Starting with the fixed value of $\omega$ = 0, Faraoni obtained
from the above equations
\begin{equation}
\alpha = \frac {1}{2} \left( 1 \pm \frac {\sqrt{3}}{\sqrt{3 + 2
\tilde {\omega}}} \right)
\end{equation}
which gives
as $\alpha \rightarrow \frac {1}{2} $ , $\tilde{\omega}$
$ \rightarrow \infty $. Under this limit, equation (20)
gives
\begin{equation}
\tilde {\phi} \rightarrow \phi_{0} + \frac {1} {\sqrt
{\tilde {\omega} }} ln \phi (\omega)
\end{equation}
It was argued that the $\phi(\omega)$ corresponding to $\omega =
0$ does not alter in the limit $\tilde{\omega} \rightarrow \infty
$ and hence one ends up with a behavior similar to equation (6).
But this argument should be taken with care as one may choose
$\phi(\omega)$ to depend on the same parameter $\alpha$ that
appears in the conformal transformation, so that $\phi(\omega)$
changes under transformation (21) even in the limit $\alpha
\rightarrow 1/2$. A simple example will illustrate the point.
Suppose $ \phi(\omega)  \sim 1+ \frac {1-2\alpha}
{\sqrt{\omega-6\alpha (\alpha -1)}}$, Under the transformation
(21), $ \phi(\omega) \rightarrow \chi (\tilde{\omega}) \sim 1 +
\frac {1} {\sqrt {\tilde {\omega}}}. $ Then we have from equation
(23) that $ \tilde {\phi} \rightarrow \phi_{0} + O\left( \frac
{1}{\tilde{\omega} } \right) $. The inclusion of the parameter
$\alpha$ in the specific choice of $\phi$ that we made in the
above example is not unreasonable as the BD field equations admit
an equivalence class of solutions for $\phi$ with a parameter
$\alpha$ (see [8]), though $\alpha $ does not appear in the BD
action. In this sense, $\alpha$ could be interpreted as some kind
of a {\it gauge parameter}. Therefore, a solution corresponding to
the choice of a particular {\it gauge}, namely, $\alpha =0$, does
not have any special status and one is free to retain $\alpha$ in
the expression for $\phi$. The conclusion that $ \tilde {\phi}
\rightarrow \phi_{0} + O \left(\frac
{1}{\sqrt{\tilde{\omega}}}\right)$
for large $\tilde{\omega}$, thus does not {\it necessarily} follow
from equation (23).  However, if $\phi(\omega)$ is chosen not to
depend on the parameter $\alpha$ that describes the conformal
transformation, the transformed scalar field $\tilde{\phi}
(\tilde{\omega})$ will truly behave like equation (6) in accordance
with the claim of Ref. [8]. \\

We argue that the functional $\omega$ independence of the Ricci
Scalar $R$ is not {\it generally} true. Consequently, the
dependence of the BD scalar field $\phi $ on the coupling constant
$\omega$ essentially remains {\it arbitrary} when $T=0$ and not
necessarily like the one expressed in equation (6). Usually one
fixes the constants appearing in the exact solutions for $\phi$
using physical considerations: In the context of
Oppenheimer-Snyder collapse in the BD theory, this point is
illustrated in the Refs. [11,12]. Also, very recently, it has been
discussed by Miyazaki [13] that the asymptotic behavior of $\phi$
could be fixed as $\phi \rightarrow O\left( \frac {1}{\omega}
\right)$ due to the presence of cosmological matter distribution
for which $T \neq 0 $ although for local matter distribution $T$
could be zero. This idea is perfectly consistent with the Machian
nature of the Brans-Dicke theory.

\end{document}